\documentclass[10pt,twocolumn,showpacs,longbibliography,amsmath,amssymb,osajnl,floatfix,superscriptaddress]{revtex4-1}

\usepackage{graphicx}
\usepackage{dcolumn}
\usepackage{bm}
\usepackage{color}
\usepackage{txfonts}
\usepackage{microtype}
\usepackage{url}
\usepackage{epstopdf}
\usepackage{indentfirst}
\usepackage{textcomp}
\usepackage{mathrsfs}
\usepackage[colorlinks,
            linkcolor=blue,
            anchorcolor=blue,
            citecolor=blue]{hyperref}

\begin{document}

\title{Tail-wave-assisted Positron Acceleration in Nonlinear Laser Plasma Wakefields}

\author{Wei-Yuan Liu}
\affiliation{Key Laboratory for Laser Plasmas (MOE), School of Physics and Astronomy, Shanghai Jiao Tong University, Shanghai, 200240, China}
\affiliation{Collaborative Innovation Center of IFSA (CICIFSA), Shanghai Jiao Tong University, Shanghai 200240, China}

\author{Xing-Long Zhu}
\affiliation{Key Laboratory for Laser Plasmas (MOE), School of Physics and Astronomy, Shanghai Jiao Tong University, Shanghai, 200240, China}
\affiliation{Collaborative Innovation Center of IFSA (CICIFSA), Shanghai Jiao Tong University, Shanghai 200240, China}
\affiliation{Tsung-Dao Lee Institute, Shanghai Jiao Tong University, Shanghai 200240, China}

\author{Min Chen}
\email{minchen@sjtu.edu.cn}
\affiliation{Key Laboratory for Laser Plasmas (MOE), School of Physics and Astronomy, Shanghai Jiao Tong University, Shanghai, 200240, China}
\affiliation{Collaborative Innovation Center of IFSA (CICIFSA), Shanghai Jiao Tong University, Shanghai 200240, China}

\author{Su-Ming Weng}
\affiliation{Key Laboratory for Laser Plasmas (MOE), School of Physics and Astronomy, Shanghai Jiao Tong University, Shanghai, 200240, China}
\affiliation{Collaborative Innovation Center of IFSA (CICIFSA), Shanghai Jiao Tong University, Shanghai 200240, China}

\author{Feng He}
\affiliation{Key Laboratory for Laser Plasmas (MOE), School of Physics and Astronomy, Shanghai Jiao Tong University, Shanghai, 200240, China}
\affiliation{Collaborative Innovation Center of IFSA (CICIFSA), Shanghai Jiao Tong University, Shanghai 200240, China}

\author{Zheng-Ming Sheng}
\affiliation{Key Laboratory for Laser Plasmas (MOE), School of Physics and Astronomy, Shanghai Jiao Tong University, Shanghai, 200240, China}
\affiliation{Collaborative Innovation Center of IFSA (CICIFSA), Shanghai Jiao Tong University, Shanghai 200240, China}
\affiliation{Tsung-Dao Lee Institute, Shanghai Jiao Tong University, Shanghai 200240, China}

\author{Jie Zhang}
\affiliation{Key Laboratory for Laser Plasmas (MOE), School of Physics and Astronomy, Shanghai Jiao Tong University, Shanghai, 200240, China}
\affiliation{Collaborative Innovation Center of IFSA (CICIFSA), Shanghai Jiao Tong University, Shanghai 200240, China}
\affiliation{Tsung-Dao Lee Institute, Shanghai Jiao Tong University, Shanghai 200240, China}

\date{\today}

\begin{abstract}
Relativistic laser wakefield acceleration is characterized by an unsurpassed accelerating gradient, which is very suitable for electron acceleration over short distances and could be a promising candidate for next-generation compact accelerators. However, using this technique for positron acceleration is still challenging because positively charged particles are naturally defocused in nonlinear wakefields. Here we propose and numerically demonstrate a scheme to accelerate an externally injected positron beam in a nonlinear laser wakefield in a regime where a tail wave is formed behind density cusps of the wakefield. This tail wave can provide a focusing force in addition to longitudinal acceleration for the positrons. Three-dimensional particle-in-cell simulations demonstrate that a trapping efficiency of positrons of nearly 100\% in the nonlinear wakefield is possible. This scheme may open a simple way for compact positron acceleration to multi-100 MeV with terawatt-class laser systems at high repetition rates without the need for special laser modes and plasma structures.
\end{abstract}

\maketitle
\section{introduction}
Plasma-based acceleration has become an active research area in recent years due to its extremely high acceleration gradient and potential application for the next generation of compact accelerators \cite{tajima1979,chen1985}. They are capable of supporting an acceleration gradient exceeding 100 GV/m, which is several orders of magnitude higher than those in conventional accelerators.
In the nonlinear regime of laser wakefield acceleration (LWFA) or plasma wakefield acceleration (PWFA), a relativistically intense drive laser or dense charge particle bunch excites a bubble-like electron cavity \cite{pukhov2002,lu2006}. This structure is inherently suitable for efficient acceleration of electrons, because electrons can be focused and accelerated simultaneously in a region covering half of the wake.
This has spurred rapid progress of wakefield acceleration and its applications in the last four decades \cite{joshi2020}. However, positrons, the antiparticle of the electrons, are generally defocused by the transverse field in the bubble of the nonlinear wakefields, making it quite difficult to achieve effective positron acceleration. Positrons at high energies of several hundered MeV to TeV are widely demanded and applied in areas ranging from nuclear medicine \cite{raichle1985}, laboratory astrophysics \cite{abeysekara2017}, to lepton colliders \cite{leemans2009}. Currently, they still rely on huge size accelerators. The realization of positron acceleration in a university-scale laboratory would be stimulating and of milestone significance in popularizing positron-based applied research.

\begin{figure}[t]		
	\includegraphics[width=0.75\linewidth]{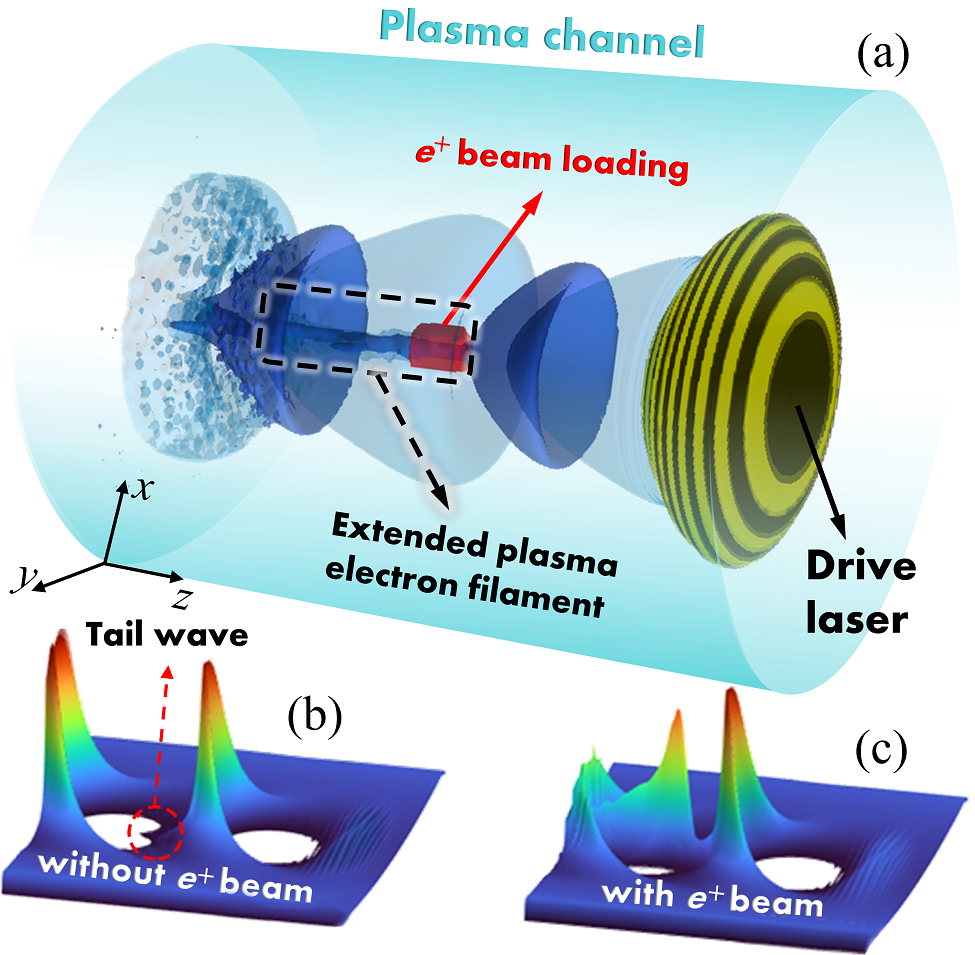}\\
	\caption{Sketch of the positron acceleration scheme. (a) A drive laser (yellow-black) and trailing positron beam (red) propagate in a parabolic plasma channel (light blue). (b) When there is no positron loaded, only a small cluster of electrons near the bubble boundary is gathered on the axis to form a tail wave. (c) When a positron beam is loaded, the tail wave is magnified with a high-density electron filament formed around the axis in the second plasma wave bucket  behind the drive pulse. }
	\label{fig1}
\end{figure}

Significant efforts have been made towards this goal in both PWFA and LWFA.
Recently, a hollow electron beam \cite{jain2015} or a positron beam \cite{corde2015,doche2017} as the driver and an electron beam in a finite-radius plasma column \cite{diederichs2019} were proposed to form an on-axis filament of plasma electrons to create a broad region in the wake cavity that focuses and accelerates positrons simultaneously. Besides, a wake excited by an electron beam in a hollow plasma channel was proposed for positron transport under zero defocusing force inside the channel \cite{gessner2016,zhou2021,silva2021}.
Despite these advances, these methods still rely on tens of GeV dense particle beams which are only accessible in few large conventional accelerator facilities. By contrast, LWFA is appealing due to its compact size, low cost and flexible beam time. In LWFA, a Laguerre-Gaussian (LG) laser pulse is suggested to drive the self-injection of the donut-like electrons and form the positron focusing region \cite{vieira2014}. Although this scheme is quite promising, it requires the high-intensity LG lasers, which is more difficult to achieve than usual Gaussian lasers. Thus, new configurations for positron acceleration in wakefields driven by a usual Gaussian laser is highly desirable.

In this work, we propose and numerically demonstrate that an extended region favorable for positron focusing and acceleration can be created in the nonlinear wakefield driven by a laser pulse with a transverse Gaussian profile. We show that when the laser intensity is marginally beyond the threshold for the nonlinear wakefields, where self-injection of electrons is not significant, a large number of plasma electrons near the density cusp at the rear of the first wave bucket behind the laser pulse can traverse the second bubble. They can form a tail wave other than composing the following wake, and some tail wave electrons can converge into an on-axis high-density filament in the bubble. Compared to previous schemes to accumulate electrons, the tail wave does not rely on special modulations of the laser mode and plasma structure, which greatly reduces the experimental difficulty. 
Moreover, when a positron beam is externally injected into the accelerating phase of such wakefields, more tail wave electrons near the positron beam can be dragged towards the beam, as shown in Fig.~\ref{fig1}. Such structure enables efficient focusing and acceleration of the loaded positron beam. We call this scenario as ``tail wave assisted positron acceleration'' (TWAA).

\section{simulation results and discussion}
We have simulated such an acceleration process by using three-dimensional particle-in-cell (3D-PIC) simulations with OSIRIS code \cite{fonseca2002}. A linearly polarized laser is used to drive a nonlinear wake in a preformed plasma channel.
To be convenient, we introduce a co-moving frame variables ($x, y, \xi=z-ct, t$), with ($x, y$) being the transverse coordinates ($r=\sqrt{x^2+y^2}$), $t$, $c$, and $z$ being the time, the speed of light, and the longitudinal coordinates, respectively. The spatiotemporal profile of the laser is given by ${\textbf {\textit a}}=a_0{\rm exp} (-{r^2}/{w_0^2}) {\rm sin^2}(\pi t/2\tau_0){\bf e_y}$, where $a_0$ is the peak vector potential of the laser, $w_0$ is the spot size, $\tau_0$ is the full width at half maximum (FWHM) duration.
The plasma channel has a parabolic transverse density profile 
\begin{equation}\label{1}
n_e=n_{e0}+\Delta nr^2/r_0^2
\end{equation}
where $n_{e0}$ is the on-axis electron density, $r_0$ is the channel width, and $\Delta n=(\pi r_er_0^2)^{-1}$ is the channel depth with $r_e=e^2/m_ec^2$ the classical electron radius.
The simulation box has a size of $40\lambda_0 (x)\times40\lambda_0 (y)\times 80\lambda_0 (z)$, divided into $200\times 200\times 2400$ grid cells with 2 macroparticles per cell for the plasma electrons (ions) and 8 macroparticles per cell for the positrons.

\begin{figure}[t]		
	\includegraphics[width=1\linewidth]{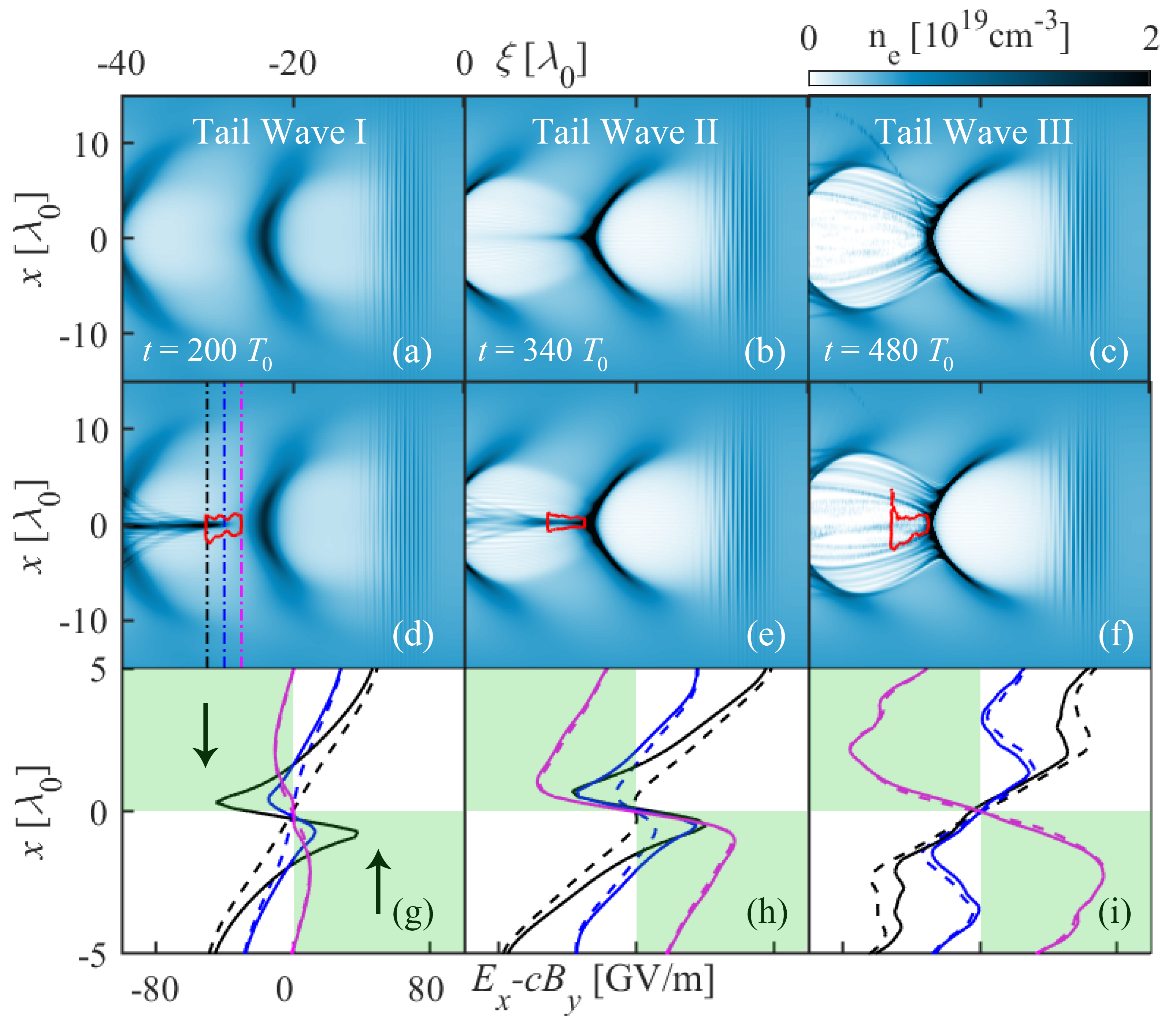}\\
	\caption{Snapshots of electron density distributions (white-blue gradients) in $\xi-x$ plane show different patterns of tail waves in unloaded wakefields [(a)-(c)] and positron loaded wakefields [(d)-(f)]. The red solid line denotes the one-tenth contour of the initial positron density. (g)-(i) Slices of transverse field $E_x-cB_y$ along $x$-axis at the back edge (black), the centroid (blue), and the front edge (magenta) of the positron beam. The results in unloaded case and the positron loaded case are denoted by the dashed lines and solid lines, respectively. The green shadings indicate the region where positrons can be transversely focused and the arrows represent the force direction.
	}
	\label{fig2}
\end{figure}

We first studied the tail wave characteristics without loading positrons.
Figure~\ref{fig2} shows the simulation results using a laser pulse with $a_0=1.6$, $w_0=12\lambda_0$, $\tau_0=7.5T_0$ ($T_0=\lambda_0/c$ is the laser period). Considering the laser wavelength $\lambda_0=1~\mu$m, it corresponds to a laser with energy of 150 mJ, peak power of 8 TW, and peak intensity of $3.5\times10^{18} {\rm W/cm^2}$.
Comparable parameters can be obtained at the commercial kHz laser systems \cite{ekspla}.
The laser pulse propagates in a plasma channel with $r_0=w_0$ and $n_{e0}\approx4\times10^{18} $ cm$^{-3}$ (giving the critical power for relativistic optical guiding $P_c\simeq5$ TW
) \cite{sprangle1992}.
Since $P>P_c$, relativistic self-focusing also contributes to the guiding of the drive laser; hence the competition between the ponderomotive potential of the driver and the radial focusing field of the wake is evolving as the laser propagates. This leads to the modification of electron trajectories behind the laser.
In response to a lower intensity drive laser, most electrons with initial transverse coordinates smaller than the laser spot size can slip into the second bubble without apparent oscillations because of the weaker plasma perturbation; and the plasma wave front where the electron density peaks is curved due to the slightly longer plasma wavelength $\lambda_{Np}\propto\sqrt{|\textbf{\textit a}|/n_e}$ on-axis than off-axis [see Fig.~\ref{fig2}(a)]. This allows a limited region behind this peak to hinder the positron defocusing [see the magenta dashed line in Fig.~\ref{fig2}(g)].
As the laser intensity increases moderately, electron accumulation and curvature at the plasma wave front are enhanced and some off-axis electrons gather at the vicinity of the central axis of the second bubble, forming a high-density filament [see Fig.~\ref{fig2}(b)]. As a consequence, the region suitable for positron focusing is expanded to approach half of a bubble [see Fig.~\ref{fig2}(h)].
At a later time, when the laser intensity increases further and electrons are transversely over-accelerated, these tail wave electrons are split transversely; thus the focusing field for positrons only exists in the axially receding and squeezed density cusp   [see Figs.~\ref{fig2}(c) and (i)] \cite{lotov2007}.

\begin{figure*}[t]		
	\includegraphics[width=0.8\linewidth]{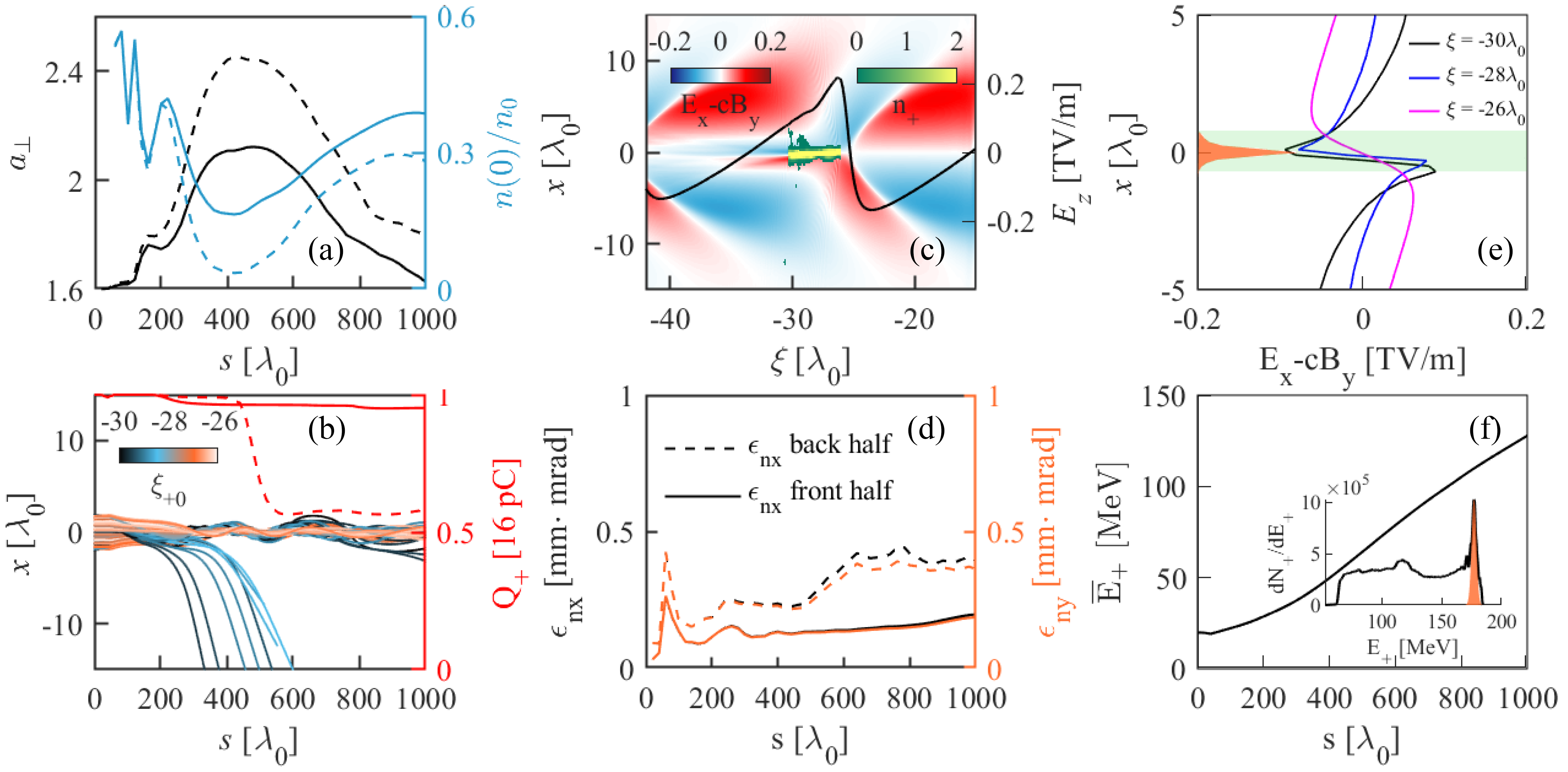}\\
	\caption{(a) Evolution of the peak laser vector potential (black) and the minimum on-axis electron density in the first bubble (blue). The solid and dashed lines denote the case with plasma channel radius $r_0=14\lambda_0$ and $r_0=12\lambda_0$, respectively.
	(b) Positron trajectories for the case with $r_0=14\lambda_0$ (colors reprsenting the initial longitudinal locations of positrons). The red lines show the evolution of the total charge of the trapped positrons in the cases with $r_0=14\lambda_0$ and $r_0=12\lambda_0$.
	(c)-(f) represent results for the case with $r_0=14\lambda_0$.
	(c) Transverse field ($E_x-cB_y$ [TV/m], blue-red gradients), positron density ($n_+$ [$10^{19}$ cm$^{-3}$], green-yellow gradients), and longitudinal field slice along $r=0$ ($E_z$, solid line) at an acceleration length of $s=500\lambda_0$.
	(d) Normalized transverse emittance $\epsilon_{nx}$ (black) and $\epsilon_{ny}$ (orange) at the back half (dashed line) and the front half (solid line) of the positron beam as a function of $s$.
	(e) Transverse fields at three different locations: $\xi=-30\lambda_0$, $\xi=-28\lambda_0$, and $\xi=-26\lambda_0$. The orange shading shows the location and the transverse shape of the positron beam, and the green shading indicates the focusing field of the beam.
	(f) Evolution of the average energy of positrons. The inset: final energy-spectrum of the beam, where the orange shading is the symmetric gaussian fit to the peak $E_+\approx177$ MeV.}
	\label{fig3}
\end{figure*}

The dynamics of tail waves can be further modified when a witness positron beam is loaded. Once a positron beam is encompassed by plasma electrons, it spontaneously attracts electrons inward to maintain the charge neutrality \cite{hogan2003}.
When the plasma density is uniform, the net focusing force of the positron beam on electrons has a linear dependence on the electron density \cite{su1990}. Similarly, for a positron beam propagating in the tail wave, the transverse sucking is positively corelated with the density of the tail wave electrons within ``reach'' of the beam.
To investigate such effects, a uniformly distributed positron beam  ($n_{+0}=0.5n_{e0}$) in a cylindrical shape (i.e. length of $L_+=4\lambda_0$ and transverse radius of $r_+=2\lambda_0$) with an initial energy of 20 MeV follows the drive laser and is placed in the second bubble, where the back end of the beam is located at $\xi_{+, l}=-30\lambda_0$ where $E_z\approx0$. Other laser and plasma parameters are the same as those mentioned above. 
As it can be seen in Figs.~\ref{fig2}(d-f), due to the relaxation time for the electrons in the off-axis tail wave to reach the positron beam \cite{lee2001}, a significant enhancement of the on-axis electron density only occurs when the tail waves are formed at stage I and II. 
Therefore, the transverse field in the front of the beam makes little difference whether positrons are loaded or unloaded, whereas in the rear of the beam it can be turned from a defocusing field to a focusing field as long as the combined electron density of the original on-axis tail wave electrons and the sucked-in electrons is high enough to exceed the background ion density [see Figs.~\ref{fig2}(g-i)].
In general, the focusing strength is proportional to the beam current, thus the elongated electron filament does not provide a uniform focusing strength along different $\xi$ \cite{hogan2003}, which leads to the inhomogeneous compression over the length of the beam [see Figs.~\ref{fig2}(d-e)]. 
Meanwhile, because positrons in this example are relativistic with $\gamma_+>\gamma_p$, where $\gamma_+$ and $\gamma_p\simeq\omega_0/\omega_p$ ($\omega_0$ is the laser frequency and $\omega_p$ is the plasma wave frequency) are the Lorentz factor of the beam and the plasma wave, respectively, they slowly move forward while the density of ambient tail wave electrons gradually increases. This is beneficial for trapping more positrons, not just the first half of the beam. When tail wave III occurs, most positrons in the rear part of the beam are accelerated transversely [see Fig.~\ref{fig2}(f)], which results in the potential loss of some accelerated positrons. Fortunately, when the laser is continuously guided in the channel, the laser intensity is decreased [see the black dashed line in Fig.~\ref{fig3}(a)] so that tail wave III can gradually revert to tail wave II and tail wave I. As a result, the on-axis electron filament in the positron region is recoverd and thus the positron loss ceases. Although about 60\% positrons are finally trapped [see the red dashed line in Fig.~\ref{fig3}(b)], it is preferable to avoid tail wave III to achieve a more stable positron acceleration. 

The blue dashed line in Fig.~\ref{fig3}(a) shows the minimum on-axis electron density in the first bucket behind the driver $n(0)/n_{e0}$ in the example, which is used to describe the blowout degree of the plasma electrons. It varies with the tail wave patterns. 
At the tail wave I and II stages, typically there are $n(0)/n_{e0}\gtrsim0.2$ and $0.1\lesssim n(0)/n_{e0}\lesssim0.2$, respectively, whereas there is $n(0)/n_{e0}\lesssim0.1$ for tail wave III.
Because $n(0)/n_{e0}\propto-a_{\perp}^2/(\sqrt{1+a_{\perp}^2}\omega_p^2w_0^2)$ \cite{esarey2009}, it is possible to keep $n(0)/n_{e0}\gtrsim0.1$ througout the wakefield generation by optimizing the initial laser-plasma parameters.
We noticed that when a light pulse propagating in a parabolic plasma channel profile, the index of refraction in the plasma channel satisfies \cite{lu2007}
\begin{equation}\label{2}
\eta\simeq1-\frac{\omega_p^2}{2\omega_0^2}(1+\frac{\Delta n}{n_0}\frac{r^2}{r_0^2}+\frac{\delta n}{n_0}-\frac{a_0^2}{8}).
\end{equation}
If one increases the channel radius and makes sure $r_0>w_0$, the effects of relativistic self-focusing will be weakened as $\eta$ increases.
For this purpose, we adopt a wider plasma channel with $r_0=14\lambda_0$ and keep other parameters the same as those in Fig.~\ref{fig2}(d).
Due to the limited increase of the laser intensity [see the black solid line in Fig.~\ref{fig3}(a)], only I and II type tail waves are excited in this case because $n(0)/n_{e0}$ is increased compared to that in the case of $r_0=12\lambda_0$ [see Fig.~\ref{fig3}(a)]. Accordingly, more efficient positron trapping is successfully achieved, where more than 95\% of the initial charge can be accelerated after a propagation distance of $s=1000\lambda_0$ [see the red solid line in Fig.~\ref{fig3}(b)]. 
The 5\% positron charge loss originates mainly from the latter middle area of the beam during the stage of tail wave I [see Fig.~\ref{fig3}(b)], which is caused by the delayed transition of the tail wave pattern due to the weakened laser focusing.
Note that the evolution time scale of positrons is the betatron period $\tau_{\beta}=2\pi\sqrt{\gamma_+/[{\rm d}E_r/{\rm d}r]_{E_r=0}}$, where ${\rm d}E_r/{\rm d}r$ is the slope of the transverse field and it varies with $\xi$. As a result, positrons with different longitudinal positions are in different phases of the oscillations for a given $s$ and the front ones have a more stable acceleration [see Fig.~\ref{fig3}(b)]. Since the focusing field is prolonged during the acceleration due to the accumulated electrons, the charge loss is limited. 

\begin{figure}[b]
	\includegraphics[width=1.0\linewidth]{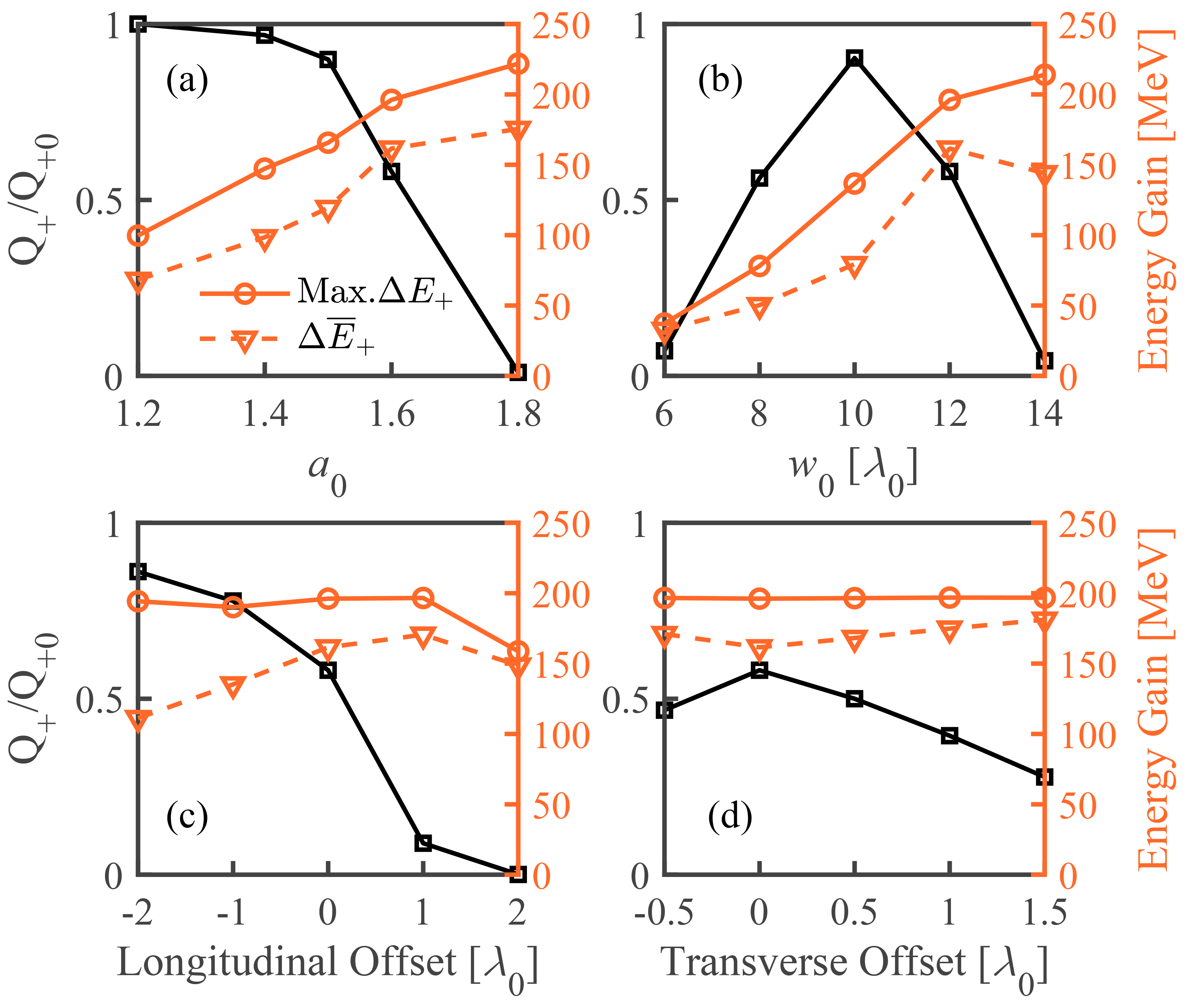}\\
	\caption{Variations of the trapping efficiency $Q_+/Q_{+0}$ (black, square markers), maximum energy gain (orange, circular markers), and average energy gain (orange, triangular markers) of the finally accelerated positrons at $s=1000\lambda_0$ with respect to the parameters of the laser $a_0$ (a), $w_0$ (b) and of the positron beam, i.e., different longitudinal (c) or transverse offsets (d) of the beam, respectively.}
	\label{fig4}
\end{figure}

To study the beam emittance evolution, we show in Figs.~\ref{fig3}(c) and (e) the details of the focusing field and the positron density at the propagation distance of $s=500\lambda_0$ where the laser intensity is the highest.  
As one can see, the focusing field for positrons is extended to a length nearly occupying the whole second wake bubble. The positron oscillation causes a slight asymmetry in the focusing field at the rear of the beam [see Fig.~\ref{fig3}(c)], and the transverse focusing region at this part is smaller [see Fig.~\ref{fig3}(e)], which leads to the formation of a beam halo, where the positrons are compressed at the center and less dense at the periphery \cite{muggli2008} [see Fig.~\ref{fig3}(c)], the transverse size of the positron beam is reduced to about one third of its original size, and the number density is increased by tens of times.
The evolution of the normalized emittance of the positron beam along different directions, i.e., $\epsilon_{nx}$ and $\epsilon_{ny}$, is illustrated in Fig.~\ref{fig3}(d).
The emittance growth of $\epsilon_{nx}$ and $\epsilon_{ny}$ are quite limited. It increases by 0.26 (0.42) in the front (back) part of the beam at the early stage. At this stage, a small number of positrons are lost. When a stronger focusing field is generated the emittance decreases by 0.17 (0.25). After that, the emittance of the front part almost stays constant, while the back part periodically grows due to the varying localized transverse fields.
To show the positron acceleration and energy spread evolution, we plot the longitudinal field of the wake in Fig.~\ref{fig3}(c). As one can see, the longitudinal wakefield generally shows a linear distribution with a quite limited uniform region at the most front of the beam where $E_z$ is close to maximum. This is due to the positron beam loading at the rising edge of the field and the beam density peaks at the right edge.
Correspondly, as shown in Fig.~\ref{fig3}(f), the acceleration gradient gradually increases as the beam moves forward, and the final energy spectrum of the positron beam is broadened, but there is a peak at 177 MeV. The average energy gain of the positron beam is $\Delta\overline{E}_+\approx110$ MeV at $s=1000\lambda_0$; the FWHM energy spread of the peak is 3\%, and the positron charge contained in this peak is about 2 pC.

We have also studied the validity of the TWAA scheme under different initial parameters.
Figures~\ref{fig4}(a-b) present the trapping efficiency and energy gain of the finally accelerated positrons where the intensity $a_0$ and the focal spot $w_0$ of the drive laser are varied, respectively. The initial position of the back end of the positron beam is $\xi_{+, l}\approx-3/2\lambda_{Np}(0)$, and other parameters are the same as those in Fig.~\ref{fig2}(a).
As aforementioned, the tail wave pattern correlates with the laser intensity. Thus, by decreasing $a_0$, the positron trapping can be enhanced, albeit at the cost of the acceleration gradient due to $E_z\propto a_{\perp}^2/\sqrt{1+a_{\perp}^2}$, as shown in Fig.~\ref{fig4}(a). When $a_0<1.6$, more than half of the positrons can be trapped and accelerated with the acceleration gradient $G\sim100$ GV/m. When $a_0=1.8$, the laser is focused to $a_{\perp}\approx3$ and an electron-free cavity is formed such that positrons are almost evacuated before tail wave II is recovered.
The positron trapping can also be enhanced by lowering $w_0$ to the region of 8$\lambda_0$ to 12$\lambda_0$, as shown in Fig.~\ref{fig4}(b). This is attributed to the weakening of the laser focusing by a wider plasma channel, because $\Delta n_c\propto w_0^{-2}$ which leads to the increases of $\eta$. The acceleration gradient, depending on the laser focusing, has a tendency to increase with $w_0$. Note that when $w_0$ is over-reduced, the laser is quickly diffracted, so the positron trapping is restricted.
The simulation results for different longitudinal or transverse offsets of the initial positron beam position are also illustrated in Figs.~\ref{fig4}(c-d) relative to the positron presented in Fig.~\ref{fig2}(a). Longitudinally, the earlier loading within 2$\lambda_0$ has no influence on the maximum energy gain, but the trapping ratio is increased and the average energy gain decreases. In contrast, if the positron loading is delayed, the trapping ratio decreases rapidly. This is because the further away from the density spike of the plasma wave front, the lower the electron density. Transversely, the trapping tolerance has a range of about 0.5$\lambda_0$. Within this range, there is less positron loss resulted from the asymmetric focusing field, which is generated during dragging the positrons from off-axis to the axis $r=0$ by the density cusp at the rear of the first bubble.

\section{conclusion}
In conclusion, we have proposed and numerically demonstrated for the first time that positrons can be well accelerated and focused in the tail wave formed behind the density cusp of a nonlinear wakefield driven by an usual Gaussian laser pulse. The trapping efficiency can be close to 100\% when the self-focusing effect of the drive laser is weakened by using a plasma channel with a relatively large radius. This TWAA scheme is viable in a certain range of parameters, which shows its feasibility for realistic operation. In further, since the drive laser required in this regime is only at the TW level, it may be realized with high repetition with kHz laser systems, which makes it promising for the wide applications of such hundreds of MeV positron beams.

\section*{acknowledgments}
This work was supported by the National Natural Science Foundation of China (Grants No. 11991074 and 12135009) and the National Postdoctoral Program for Innovative Talents (Grant No. BX20220206). The authors would like to thank the Supercomputer Center at SJTU for providing computing resources.

\bibliography{mybib}

\end{document}